\documentclass[12pt]{article}
\newcommand{\pa}{\partial}
\newcommand{\be}{\begin{equation}}
\newcommand{\ee}{\end{equation}}
\newcommand{\bea}{\begin{eqnarray}}
\newcommand{\eea}{\end{eqnarray}}
\newcommand{\nn}{\nonumber}

\newcommand{\s}{\sigma}

\newcommand{\ep}{{\displaystyle \epsilon_{\IR}}}
\newcommand{\vep}{{\displaystyle \epsilon_{\UV}}}
\newcommand{\m}{\mu}

\newcommand{\cN}{{\cal N}}

\newcommand{\bt}[1]{{\bar t}}

\newcommand \vev [1] {\langle{#1}\rangle}
\newcommand\lr[1]{{\left({#1}\right)}}
\newcommand{\IR}{{\scriptscriptstyle \rm IR}}
\newcommand{\UV}{{\scriptscriptstyle \rm UV}}

\input epsf
\usepackage{amsmath,amsfonts,epsfig,color,latexsym}
\usepackage{amssymb,psfrag}

\renewcommand{\s}{\sigma}

\begin{document}

\thispagestyle{empty}
\null\vskip-12pt \hfill  LAPTH-1192/07 \\
\null\vskip-12pt \hfill LPT--Orsay--07--46
\vskip2.2truecm
\begin{center}
\vskip 0.2truecm {\Large\bf
{\Large Conformal properties of four-gluon \\[2mm] planar amplitudes and Wilson loops}
}\\
\vskip 1truecm
{\bf J.M. Drummond$^{*}$, G.P. Korchemsky$^{**}$ and E. Sokatchev$^{*}$ \\
}

\vskip 0.4truecm
$^{*}$ {\it Laboratoire d'Annecy-le-Vieux de
Physique Th\'{e}orique  LAPTH\footnote{UMR 5108 associ\'{e}e \`{a}
 l'Universit\'{e} de Savoie},\\
B.P. 110,  F-74941 Annecy-le-Vieux, France\\
\vskip .2truecm $^{**}$ {\it
Laboratoire de Physique Th\'eorique%
\footnote{Unit\'e Mixte de Recherche du CNRS (UMR 8627)},
Universit\'e de Paris XI, \\
F-91405 Orsay Cedex, France
                       }
  } \\
\end{center}

\vskip 1truecm 
\centerline{\bf Abstract} \normalsize \noindent We present further evidence for a
dual conformal symmetry in the four-gluon {planar} scattering amplitude in
$\cN=4$ SYM. We show that all the momentum integrals appearing in the
perturbative on-shell calculations up to four loops are dual to true conformal
integrals, well defined off shell. Assuming that the complete off-shell amplitude
has this dual conformal symmetry and using the basic properties of factorization
of infrared divergences, we derive the special form of the finite remainder
previously found at weak coupling and recently reproduced at strong coupling by
AdS/CFT. We show that the same finite term appears in a weak coupling
calculation of a Wilson loop whose contour consists of four light-like segments
associated with the gluon momenta. We also demonstrate that, due to the special form
of  the finite remainder, the asymptotic Regge limit of the four-gluon amplitude
coincides with the exact expression evaluated for arbitrary values of the
Mandelstam variables.

\newpage
\setcounter{page}{1}\setcounter{footnote}{0}

\section{Introduction}

Gluon {and quark} scattering amplitudes have long been the subject of
numerous studies in QCD. These amplitudes have a very non-trivial structure
consisting of infrared singular and finite parts. In physical infrared
safe observables like inclusive cross-sections the former cancel in the sum of
all diagrams~\cite{Kinoshita62} while the latter produce a rather complicated
function of the kinematic variables. The infrared singular part of the scattering
amplitude has a universal structure \cite{BCM84,CSS88,C} which is intimately
related to the properties of Wilson loops \cite{KR86}. This leads to an evolution
equation for the amplitude as a function of the infrared cutoff
\cite{Sen82,BS89,KOS98} which is governed, in the planar limit, by the so-called
cusp anomalous dimension of the Wilson loop~\cite{KK94,MDS06}. This anomalous
dimension first emerged in the studies of \textit{ultraviolet} cusp
singularities of Wilson loops~\cite{P80} (see also \cite{BNS81} and references
therein) and its relation to \textit{infrared} asymptotics in gauge
theories was discovered in \cite{KR86,IKR85,KR87}. The cusp anomalous dimension is very
important in QCD since it controls the asymptotic behavior of various gauge
invariant quantities like the double-log (Sudakov) asymptotics of form factors,
the logarithmic scaling of the anomalous dimension of higher-spin operators, the
gluon Regge trajectory, etc. However, unlike the singular part, the finite part
of the gluon scattering amplitude in QCD is much more involved, being given in
terms of certain special functions of the Mandelstam variables.

Recently, a lot of attention has been paid to the problem of calculating gluon
scattering amplitudes in the context of the maximally supersymmetric Yang-Mills
theory ($\cN=4$ SYM). These amplitudes have been extensively studied in
perturbation theory where they have been constructed using state-of-the-art
unitarity cut techniques \cite{BernRozowskyYan(1997),AnastasiouBernDixonKosower(2003),bds05,bcdks06,bcjk07}.
The results of these studies concern both the divergent and finite parts of the amplitude. Although the main subject of the present paper is the finite part, we start with a brief review of the IR singularities.

\subsection{Infrared divergences in gluon scattering amplitudes}

Unlike a generic gauge theory, $\cN=4$ SYM is ultraviolet finite. Despite this UV
finiteness, the gluon scattering amplitudes are still IR divergent, even though
the singular structure is much simpler {compared to QCD}, due to the fact that the coupling does not run. As in QCD, the dependence on the IR cutoff is determined by
the cusp anomalous dimension.

\begin{figure}[htbp]
\psfrag{x}[cc][cc]{$x$}
\psfrag{theta}[cc][cc]{$\vartheta$}
\psfrag{C}[cc][cc]{$C$}
\centerline{{\epsfysize5cm \epsfbox{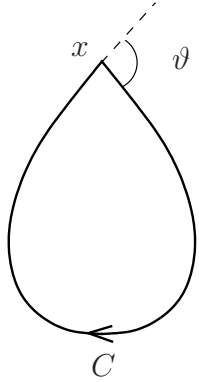}}} \caption{\small An example of a
Wilson loop with a cusp at the point $x$.}
\label{figure:cusp}
\end{figure}

The notion of cusp anomalous dimension was initially
introduced~\cite{P80,BNS81} in the context of a  Wilson loop evaluated over a
closed {(Euclidean)} contour with a cusp (see Fig.~\ref{figure:cusp}). By
definition, $\Gamma_{\rm cusp}(a,\vartheta)$ is a function of the coupling
constant $a$ and the cusp angle $\vartheta$ describing the dependence of the
Wilson loop on the {\it ultraviolet} cutoff. Later on it was realized
\cite{KR86,IKR85} that the same quantity $\Gamma_{\rm cusp}(a,\vartheta)$
determines the {\it infrared} asymptotics of scattering amplitudes in gauge
theories, for which a dual Wilson loop is introduced with an integration contour
$C$ uniquely defined by the particle momenta. The cusp angle $\vartheta$ is
related to the scattering angles and it takes large values in Minkowski space,
$|\vartheta| \gg 1$. In this limit, $\Gamma_{\rm cusp}(a,\vartheta)$ scales
linearly in $\vartheta$ to all loops \cite{KR87}
\be\label{cusp-as}
\Gamma_{\rm cusp}(a,\vartheta) = \vartheta\,\Gamma_{\rm cusp}(a) +
O(\vartheta^0)\,,
\ee
where $\Gamma_{\rm cusp}(a)$ is a function of the coupling constant only. In what
follows we shall use the term {\it cusp anomalous dimension} in this restricted
sense,  to denote the quantity  $\Gamma_{\rm cusp}(a)$. In a dimensionally
regularized four-gluon scattering amplitude, {the IR poles exponentiate and}
$\Gamma_{\rm cusp}(a)$ {controls} the coefficient of the double pole {in
the exponent}.

{The two-loop expression for $\Gamma_{\rm cusp}(a,\vartheta)$ and
$\Gamma_{\rm cusp}(a)$ in a generic (supersymmetric) Yang-Mills theory can be
found in \cite{KR87,BGK03}}. Among the important results of {the recent studies
in $\mathcal{N}=4$ SYM} was the calculation of $\Gamma_{\rm cusp}(a)$ at three
\cite{bds05} and at four \cite{bcdks06,cachazo} loops. The three-loop value of
$\Gamma_{\rm cusp}(a)$ confirmed the maximal transcendentality conjecture of
\cite{KLOV} based on advanced calculations of twist-two anomalous dimensions in
QCD \cite{moch}. The four-loop value provided support for a conjecture
\cite{bes06} about the form of the all-loop cusp anomalous dimension derived from
Bethe Ansatz equations.

\subsection{Finite part of the four-gluon amplitude}

As mentioned before, the finite part of scattering amplitudes in QCD is a very complicated object. The situation turns out to be radically simpler in $\cN=4$ SYM. One of the main results of Bern {\it et al} is a very interesting all-loop iteration
conjecture about the IR finite part of the color-ordered planar amplitude which
takes the surprisingly simple form
\begin{equation}\label{simpleform}
    \ln {\cal M}_4  = \mbox{[IR divergences]} + \frac{\Gamma_{\rm cusp}(a)}{4} \ln^2 \frac{s}{t} + \mbox{const}\ ,
\end{equation}
where $a=g^2N/(8\pi^2)$ is the coupling constant and $s$ and $t$ are the
Mandelstam kinematic variables.\footnote{To avoid the appearance of an imaginary
part in $\ln {\cal M}_4$, it is convenient to choose $s$ and $t$ negative.}
Compared to QCD, we see the following important simplifications: (i) the
complicated functions of $s/t$ appearing in QCD expressions are replaced by the
elementary function $\ln^2(s/t)$; (ii) no higher powers of logs appear in $\ln
{\cal M}_4$ at higher loops; (iii) the coefficient of  $\ln^2(s/t)$ is determined
by $\Gamma_{\rm cusp}(a)$, just like the coefficient of the double log in the IR
divergent part. This conjecture has been verified up to three loops for
four-gluon amplitudes in \cite{bds05} (a similar conjecture for $n$-gluon
amplitudes \cite{bds05} has been confirmed for $n=5$ at two loops in
\cite{Cachazo:2006tj,5point}).

Recently Alday and Maldacena  \cite{am07} have proposed a prescription for
studying gluon scattering amplitudes at strong coupling via AdS/CFT. Their
calculation produced exactly the same form (\ref{simpleform}) of the finite part,
with the strong-coupling value of $\Gamma_{\rm cusp}(a)$ obtained from the
semiclassical analysis of \cite{GPK}. This result constitutes a non-trivial test
of the AdS/CFT correspondence. Nevertheless, on the gauge theory side the deep
reason for the drastic simplification of the finite part remains unclear.

In this paper we present arguments that the specific form of the finite part of
the four-gluon amplitude, both in perturbation theory and at strong coupling, may
be related to a hidden conformal symmetry of the amplitude. To avoid
misunderstandings, we wish to stress that this is not a manifestation (at least,
not in an obvious way) of the underlying conformal symmetry of the  $\cN=4$ SYM
theory.

\subsection{Dual conformal symmetry} \label{dual}

The starting point of our discussion is the on-shell { planar} four-gluon scattering
amplitude given in terms of IR divergent momentum scalar-like integrals,
regularized dimensionally by going to $D=4-2\ep$ ($\ep<0$) dimensions. These
integrals possess a surprising symmetry which we shall refer to as {\it dual
conformal symmetry}. Its presence was first revealed, up to three loops, in
\cite{magic} and  was later on confirmed at four loops in \cite{bcdks06}.\footnote{The four-loop amplitude was constructed in \cite{bcdks06} without any assumption about dual conformal invariance. However, it turned out that only dual conformal integrals appear in it.}  Also,
it was recently used in \cite{bcjk07} as a guiding
principle to construct the five-loop
amplitude.

The central observation of \cite{magic} is that in order to uncover this dual
conformal symmetry of the on-shell integrals, one has to go through three steps:
(i) assign  `off-shellness' (or `virtuality') to the external momenta entering
the integrand; (ii) set $\ep=0$; (iii) make the change of variables
\begin{equation}\label{chva}
    p_i = x_i - x_{i+1}\ .
\end{equation}
In this way the integrals cease to diverge because the virtualities of the
external momenta serve as an IR cutoff. The resulting four-dimensional integrals
become manifestly conformal in the dual space description with `coordinates'
$x_i$. Here we stress that these are not the original coordinates in position
space (the Fourier counterparts of the momenta), but represent the momenta
themselves. The conformal symmetry is then easily seen by doing conformal
inversion $x^\mu \,\to \, x^\mu/x^2$ and counting the conformal weights at the
integration points.

The authors of \cite{bcdks06} and  \cite{bcjk07} have  noticed
that some of the dual conformal four- and five-loop integrals that they could list, in reality
do not contribute to the amplitude. In the present paper we give the explanation
of this fact, namely, all the non-contributing integrals are in fact divergent
even off shell.

It is important to realize that taking off shell the integrals which appear
in the on-shell calculations of Bern {\it et al}, in order to reveal their
dual conformal properties, does not mean that we know the exact form of the
off-shell amplitude regularized by the virtuality of the external legs. Indeed,
there are indications that the complete off-shell amplitude may involve more
integrals which vanish in the on-shell regime (see the discussion in Section
\ref{disc}). Nevertheless, inspired by the strong evidence for a conformal
structure from the perturbative calculations of Bern {\it et al}, we make the
conjecture that the full off-shell amplitude is conformal. Then, combining this
assumption with the basic properties of factorization of four-gluon amplitudes
into form factors and the exponentiation of the latter (valid also in the
off-shell regime), we deduce that the special form of the finite remainder
discussed above is a direct consequence of the dual conformal invariance. This is
one of the main points of the present paper.

\subsection{Light-like Wilson loops}

Another point we would like to make concerns the recent proposal of Alday and
Maldacena for a string dual to the four-gluon amplitudes \cite{am07}. They
identify $\ln {\cal M}_4$ with the area of the world-sheet of a classical string
in AdS space, whose boundary conditions are determined by the gluon momenta.
Quite interestingly, they use exactly the same change of variables (\ref{chva})
which they interpret as a T-duality transformation on the string world-sheet. Then they apply conformal transformations in the dual space to relate different
solutions to the string equations of motion. Remarkably, their calculation looks very similar to that of the expectation value of a Wilson loop made out
of four light-like segments $(x_i,\ x_{i+1})$ in $\mathcal{N}=4$ SYM at strong
coupling~\cite{M98,Kr02}.

Motivated by these findings, in the present paper
we revisit the calculation of such a light-like Wilson loop at weak coupling for
generic values of $s$ and $t$. We establish the correspondence between the IR
singularities of the four-gluon amplitude and the UV singularities of the Wilson
loop and extract the finite part of the latter at one loop. Remarkably, our
result is again of the form (\ref{simpleform}). This indicates that the duality
between gluon amplitudes and Wilson loops discussed by Alday and Maldacena is
also valid at weak coupling.

The relationship between gluon scattering amplitudes and light-like Wilson loops
has already been investigated at weak coupling in QCD in the context
of the Regge limit $s\gg -t>0$ \cite{KK96}. Here we examine the asymptotic behavior of the
$\cN=4$ SYM four-gluon amplitude in the Regge regime and demonstrate
that, due to the special form of the finite remainder, the amplitude is Regge
exact. This means that the contribution of the gluon Regge trajectory to the
amplitude coincides with its exact expression evaluated for arbitrary values of
$s$ and $t$. This property is in sharp contrast with QCD, where the simplicity of
the Regge limit is lost if the amplitude is considered in a general regime.

\section{Perturbative structure of planar four-gluon\\ scattering amplitudes. Evidence for off-shell\\ conformal symmetry}

In this section we discuss some properties of the loop integrals appearing in the
perturbative calculation of the planar four-gluon scattering amplitude. It is
expressed in terms of dimensionally regularized Feynman integrals with the
external legs on shell. However, if one takes the legs off shell and restricts
the integrals to four dimensions, they exhibit an unexpected dual conformal
symmetry.

\begin{figure}[htbp]
\centerline{{\epsfysize5cm \epsfbox{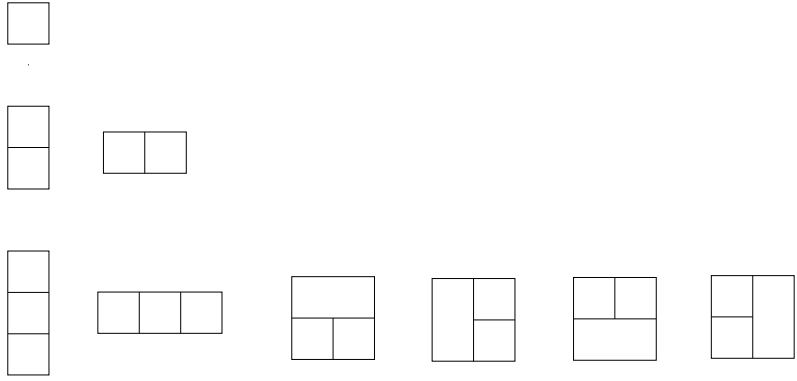}}}
\caption{\small Integral topologies up to three loops}
\label{figure:4ptamp}
\end{figure}

The planar four-gluon amplitude in $\cN=4$ SYM at one loop \cite{Brink,Grisaru}
and two loops \cite{BernRozowskyYan(1997),AnastasiouBernDixonKosower(2003)} is
expressed in terms of ladder (or scalar box) integrals. Such integrals have long
been known to have special properties. In particular, when treated off shell in
four dimensions, they are conformally covariant \cite{Broadhurst}. At three loops
\cite{bds05} the amplitude is given by three-loop ladder integrals as well as one
new type of integral, the so-called `tennis court' (see Fig. \ref{figure:4ptamp}). In \cite{magic} it was shown
that the `tennis court' integral is also conformally covariant.

\begin{figure}[htbp]
\psfrag{x1}[cc][cc]{$x_1$} \psfrag{x2}[cc][cc]{$x_2$} \psfrag{x3}[cc][cc]{$x_3$}
\psfrag{x4}[cc][cc]{$x_4$} \psfrag{p1}[cc][cc]{$p_1$} \psfrag{p2}[cc][cc]{$p_2$}
\psfrag{p3}[cc][cc]{$p_3$} \psfrag{p4}[cc][cc]{$p_4$} \psfrag{x5}[cc][cc]{$x_5$}
\centerline{{\epsfysize5cm \epsfbox{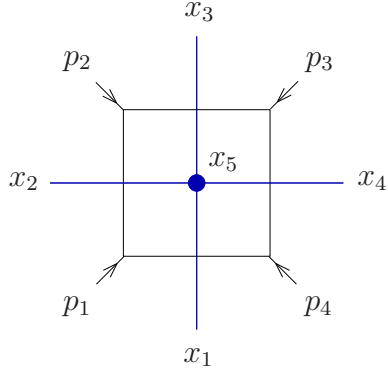}}} \caption{\small  Dual diagram
for the one-loop box}
%
\label{figure:dualdiag}
\end{figure}

As an example, consider the  one-loop scalar box integral
\begin{equation}\label{1box}
    I^{(1)} = \int \frac{d^Dk}{k^2(k-p_1)^2(k-p_1-p_2)^2(k+p_4)^2}\ .
\end{equation}
It is a function of four gluon momenta $p_i$ such that $\sum_{i=1}^4 p_i=0$. When
the external legs are put on shell, $p^2_i=0$, the integral becomes infrared
divergent and needs to be regularized. One way to do this is to change the
dimension from $D=4$ to $D=4-2\ep$, $\ep<0$ (dimensional regularization). Another
way is to have the external legs slightly off shell, $p^2_i \neq 0$ and later on
to take the limit $p_i^2\to 0$. In the latter approach we can keep $D=4$ which
allows us to reveal the conformal properties of the integral. This is done by
introducing a dual `coordinate' description (see Fig. \ref{figure:dualdiag}).

We define dual variables $x_i$ by (with $x_{jk}\equiv x_j-x_k$)
\begin{equation}\label{defvaria}
p_1 = x_{12}\,,\quad  p_2 = x_{23}\,, \quad p_3 = x_{34} \,, \quad p_4 =
x_{41}\,,\quad k=x_{15} \,,
\end{equation}
so that $\sum_i p_i=0$.  We stress that these are not the coordinates in the
original position space (the Fourier counterparts of the momenta), but simply a
reparametrization of the momenta (note the `wrong' dimension of mass of $x_{i,
i+1}$). In terms of these new variables the integral (\ref{1box}) takes the form
\begin{equation}
I^{(1)}=\int \frac{d^D x_5}{x_{15}^2 x_{25}^2 x_{35}^2 x_{45}^2} \ .
\end{equation}
It is manifestly invariant under translations and rotations of the $x$
coordinates. It is also covariant under conformal inversion,
\begin{equation}
x^\m \longrightarrow \frac{x^\m}{x^2}\ : \qquad \frac{1}{x^2_{ij}} \longrightarrow \frac{x^2_i x^2_j}{x^2_{ij}}  \ , \qquad d^Dx \longrightarrow \frac{d^Dx}{(x^2)^D}\ ,
\end{equation}
provided that the transformation of the propagators at the integration point
$x_5$ is exactly compensated by the transformation of the measure. This can only
happen if $D=4$. Then the integral is equal to a conformally covariant factor
multiplied by a function of the conformally invariant cross-ratios $u$ and
$v$,\footnote{We denote the cross-ratios by $u$ and $v$ and reserve $s$ and $t$
to denote the Mandelstam variables.}
\begin{equation}\label{cross}
u=\frac{x_{12}^2 x_{34}^2}{x_{13}^2 x_{24}^2}\ , \hspace{40pt} v=\frac{x_{14}^2x_{23}^2}{x_{13}^2 x_{24}^2}\ .
\end{equation}
Thus we have
\begin{equation}\label{onebox}
I^{(1)} = \frac{i\pi^2}{x_{13}^2 x_{24}^2} \Phi^{(1)}(u,v)\ ,
\end{equation}
where the function $\Phi^{(1)}$ is expressed in terms of logs and dilogs \cite{Davydychev}.

We wish to point out that this conformal invariance of the loop integrals is {\it
a priori} unrelated (or at least not related in an obvious way) to the conformal symmetry
of the underlying $\cN=4$ SYM theory. For this reason we prefer to call this
property of the loop integrals `dual conformal invariance'.

\begin{figure}[htbp]
\psfrag{1}[cc][cc]{$1$}
\psfrag{2}[cc][cc]{$2$}
\psfrag{3}[cc][cc]{$3$}
\psfrag{4}[cc][cc]{$4$}
\psfrag{5}[cc][cc]{$5$}
\psfrag{6}[cc][cc]{$6$}
\psfrag{7}[cc][cc]{$7$}
\centerline{{\epsfysize4cm \epsfbox{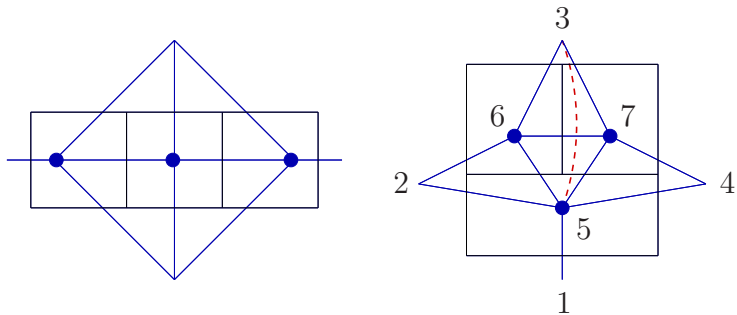}}}
\caption{\small Dual diagrams for the three-loop box and for the `tennis court'
with its numerator} \label{figure:3ladandtc}
\end{figure}

The conformal properties of the higher-order scalar box integrals and of the `tennis court' can
be seen in the same way. For the `tennis court' there is the new feature that the
integrand contains a numerator (denoted by a dashed line). It is a positive power
of $x_{35}^2$ connecting the external point 3 to the integration point 5 where more than four
propagators join together (Fig. \ref{figure:3ladandtc}). The r\^ole of this numerator is to compensate the surplus
of conformal weight due to the propagators at this internal point, thus
maintaining conformal covariance.

At four loops the conformal pattern continues. The amplitude was constructed in \cite{bcdks06} and it is again expressed entirely in terms of dual conformal integrals. The authors of \cite{bcdks06} give a list of ten such planar integrals which are non-vanishing on shell and have only log (no power) singularities. All of them satisfy the formal requirement of dual conformal covariance, namely that the conformal weight of the integrand at each integration point should be four. However, it turns out that of the ten integrals, eight contribute to the amplitude and two do not. We are now able to give a simple explanation: By inspection one can see that the two non-contributing integrals are in fact divergent in four dimensions (even off shell). Thus they do not have well-defined conformal properties, since they require a regulator to exist at all. Furthermore, all the integrals which {do} contribute to the amplitude are finite and hence conformal in four dimensions.

\begin{figure}[htbp]
\psfrag{integral}[cc][cc]{$\quad \displaystyle \sim \int \frac{\rho^{15}d\rho}{\rho^{16}}$}
\centerline{{\epsfysize8cm \epsfbox{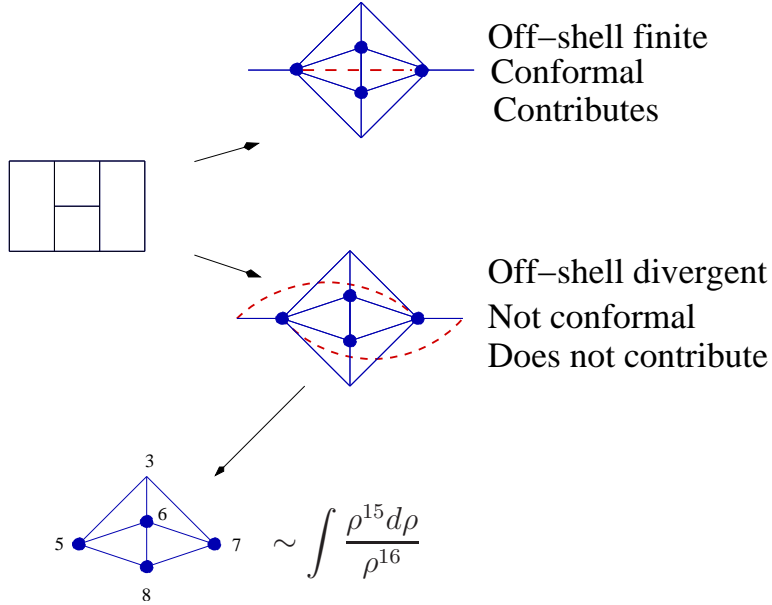}}}  \caption{\small Examples
of off-shell finite and divergent integrals with the same topology but different
numerators} \label{figure:4loopexample}
\end{figure}

We illustrate the nature of the divergence in Fig. \ref{figure:4loopexample}. The
integrals in this figure correspond to the integrals (d) and (d') of
\cite{bcdks06}. They differ only in the distribution of the numerator factors.
The first integral is finite off shell in four dimensions and contributes to the
amplitude. The second integral contains the four-loop structure indicated at the
bottom of the diagram. When all four integration points $x_{5,6,7,8}$ approach an
external point, e.g. $x_3$, connected by three propagators, {the integral scales
at short distances $\rho^2=x_{53}^2+x_{63}^2 + x_{73}^2 + x_{83}^2\to 0$ as%
\footnote{Here it is tacitly assumed that for $s,\, t$ negative and with the external
legs off-shell, the Feynman integrals can be analytically continued to the
Euclidean region. }
\be
\int \frac{ d^4 x_5 d^4 x_6 d^4 x_7 d^4 x_8}{x_{53}^2 x_{63}^2 x_{73}^2 x_{56}^2
x_{67}^2 x_{78}^2 x_{58}^2 x_{68}^2}\sim \int \frac{\rho^{15}d\rho}{\rho^{16}}
\ee
and therefore it is divergent as $\rho\to 0$}. The first integral does not suffer
from this divergence due to the different numerator structure which softens the
behaviour in the equivalent region of integration by one power of $\rho^2$.

In principle, the fact that some integrals are divergent off shell in four dimensions is not a problem in the study of dimensionally regularized on-shell amplitudes. It is, however, striking that an integral contributes to the amplitude if and only if it has well-defined conformal properties off shell. The pattern also continues to five loops. In \cite{bcjk07} the five-loop amplitude was constructed based on the assumption that it be a linear combination of dual conformal integrals. The authors found 59 such integrals of which only 34 contribute to the amplitude. Once again, all 34 contributing integrals are truly conformal, i.e. they are finite off shell in four dimensions, and the 25 non-contributing integrals are divergent. The divergences can be either of the same form as above (i.e. a four-loop subdivergence) or of the type where all five integration points approach an external point.

Another remarkable property of the amplitude up to five loops is that the dual
conformal integrals all come with coefficient $\pm 1$. It seems reasonable to
conjecture that these properties hold to all loops, i.e. that the all-order
planar four-gluon amplitude has the form (after dividing by the tree amplitude)
\footnote{Here we are using the same conventions as \cite{bds05}.},
\begin{equation}
\mathcal{M}_4 = 1+\sum_\mathcal{I} \s(\mathcal{I})\ a^{l(\mathcal{I})}
\mathcal{I}\,,
\label{all-loop}
\end{equation}
where the sum runs over all true dual conformal integrals, $\mathcal{I}$. In this
formula, $l(\mathcal{I})$ is the loop order of the integral, $a$ is the coupling
and $\s(\mathcal{I})=\pm 1$. The sign can be determined graph by graph by using
the unitarity cut method as described in \cite{bcjk07} but a simple rule for it
is still lacking.

Thus we have seen that there is very strong evidence for a dual conformal
structure behind the planar four-gluon amplitude to all orders. Here we wish to
stress once more that knowing the on-shell amplitudes as given in terms of
dimensionally regularized integrals and simply changing the regulator does not
mean that we have obtained the full off-shell amplitude. The latter may involve
further integrals. At the present stage we can only conjecture that
they are also conformal up to terms which vanish with the removal of
the infrared cutoff.

\section{Factorization and exponentiation of infrared\\ singularities. Finite part and dual conformal\\ invariance}\label{consect}

In this section we analyze the general structure of infrared divergences of
four-gluon scattering amplitudes in the on-shell and off-shell regimes. In both cases, they
factorize in the planar limit into a product of form factors in the $s-$ and
$t-$channels. The latter are known to have a simple exponential form governed by
the cusp anomalous dimension and two other (subleading) anomalous
dimensions \cite{on-shell,Korchemsky2}. This completely determines the infrared
divergent part of the four-gluon amplitude, but leaves the finite part arbitrary.
We show that the simple requirement of off-shell dual conformal invariance fixes
the finite part to exactly the form (\ref{simpleform}) observed at weak \cite{bds05} and strong
\cite{am07} coupling.

\subsection{One-loop example}\label{onelex}

Let us first consider the one-loop four-gluon amplitude (divided by the tree
amplitude) given by the one-loop box integral $I^{(1)}$ defined in
(\ref{1box})\,:
\begin{equation}\label{1loopamp}
    \mathcal{M}_4 = 1 + a M^{(1)} + O(a^2)= 1 -\frac{a }{2} st I^{(1)} + O(a^2)\ ,
\end{equation}
where the coupling is given by
\begin{equation}\label{cou}
    a = \frac{g^2N}{8 \pi^2}
\end{equation}
and $s=(p_1+p_2)^2$ and $t=(p_2+p_3)^2$ are the Mandelstam variables. In the on-shell regime
($p^2_i=0$) we can use the dimensional regularization scheme where the integral
is multiplied by a normalization factor including  the regularization scale
$\mu$. Expanding the integral in powers of the regulator $\ep$ one finds
\cite{bds05}
\begin{equation}\label{1lexpa}
M^{(1)}_{ \rm on-shell} =
\left(\frac{\mu_{\IR}^2}{-s}\right)^{\ep}\left[-\frac{2}{\epsilon_\IR^2} -
\frac{1}{\ep} \ln \frac{s}{t} + 4 \zeta_2 + O(\ep)\right]\
    ,
\end{equation}
where $\mu_{\IR}^2$ is related to the dimensional regularization scale as
\be
\mu_{\IR}^2 = 4\pi {\rm e}^{-\gamma}\mu^2
\ee
and $\gamma$ is the Euler constant. This amplitude can be split into a
divergent and a finite part in such a way that the latter does not depend on the
scale $\mu_{\IR}$:
\begin{equation}\label{M1-sum}
M^{(1)}_{\rm on-shell} = {D}^{(1)}_{\rm on-shell} + F^{(1)}_{\rm on-shell} + O(\ep)\ ,
\end{equation}
with the divergent part given by
\begin{equation}\label{divepart}
{D}^{(1)}_{\rm on-shell} = -\frac{1}{\epsilon_\IR^2} \left[ \left(
\frac{\mu_{\IR}^2}{-s}\right)^\ep + \left(\frac{\mu_{\IR}^2}{-t}\right)^\ep
\right]\ ,
\end{equation}
and the finite part given by
\begin{equation}\label{fponsh}
F^{(1)}_{\rm on-shell} = \frac{1}{2} \ln^2 \frac{s}{t} + 4 \zeta_2\ .
\end{equation}

{As mentioned in the introduction, the dependence of the on-shell scattering
amplitude $\mathcal{M}_4^{\rm on-shell}$ on the IR cutoff $\mu_{\IR}^2$ is
governed by an evolution equation. In $\mathcal{N}=4$ SYM and in the planar
limit, this equation takes the simple form~\cite{on-shell}
\be\label{defcusp}
\left(\frac{\pa}{\pa \ln \mu_{\IR}^2}  \right)^2 \ln\mathcal{M}_4^{\rm on-shell}
= -\Gamma_{\rm cusp}(a) + O(\ep)\,,
\ee
where $\Gamma_{\rm cusp}(a)$ is the cusp anomalous dimension of a Wilson loop,
Eq.~(\ref{cusp-as}). Thus, substituting ${\cal M}_4^{\rm
on-shell}=1+a M_{\rm on-shell}^{(1)}$ into (\ref{defcusp}) we obtain
$\Gamma_{\rm cusp}(a)=2a +
O(a^2)$.\footnote{Note that the so-called soft anomalous dimension $\gamma_K$
\cite{C,Kodaira81} is related to $\Gamma_{\rm cusp}$ as follows:
$\gamma_K=2\Gamma_{\rm cusp}$.}}

Let us now redo the same calculation, but keeping the integral (\ref{1box}) in
four dimension and using instead a small `virtuality' of the external legs
$p^2_i=-m^2$ as an infrared cutoff. The off-shell amplitude has been calculated
in \cite{Grisaru} and is again given by the one-loop box integral as in
(\ref{1loopamp}). This time we can use the conformal expression (\ref{onebox}) of
the integral with the function $\Phi^{(1)}(u,v)$ from \cite{Davydychev}.
According to (\ref{defvaria}) and (\ref{cross}), the conformal cross-ratios $u$
and $v$ are now given by\footnote{We are not obliged to set all virtualities
$p^2_i$ equal and we do it here only for simplicity. Keeping  $p^2_i$ independent
would allow us to examine the full off-shell conformal structure. }
\begin{equation}\label{crossoffsh}
u=\frac{x_{12}^2x_{34}^2}{x_{13}^2x_{24}^2} = \frac{p_1^2 p_3^2}{st} = \frac{m^4}{st}\ ,
\hspace{40pt}
v=\frac{x_{14}^2 x_{23}^2}{x_{13}^2 x_{24}^2} = \frac{p_4^2 p_2^2}{st}=\frac{m^4}{st}\ .
\end{equation}
It is then not hard to find the expansion of the integral in terms of the cutoff $m$:
\begin{equation}\label{amplonsh}
    M^{(1)}_{\rm off-shell} = -\frac{1}{2} \ln^2 \left(\frac{m^4}{st} \right) - \zeta_2 + O(m)\ .
\end{equation}
It can again be split into a divergent and a finite part as follows:
\begin{equation}\label{splitoff}
    M^{(1)}_{\rm off-shell} = {D}^{(1)}_{\rm off-shell} + F^{(1)}_{\rm off-shell} + O(m)\ ,
\end{equation}
with
\begin{equation}\label{divepartoff}
{D}^{(1)}_{\rm off-shell} = -\ln^2\left( \frac{m^2}{-s}\right) - \ln^2\left( \frac{m^2}{-t}\right)\ ,
\end{equation}
and
\begin{equation}\label{finiteoff}
F^{(1)}_{\rm off-shell} = \frac{1}{2} \ln^2 \frac{s}{t} - \zeta_2\ .
\end{equation}
Notice that the double-pole singularity of the dimensionally regularized
amplitude (\ref{divepart})  has been replaced by a double-log (log-squared)
singularity in the cutoff $m$. As before, the finite part (\ref{finiteoff}) does
not depend on the IR cutoff. We remark that the finite part is the same in both
schemes, except for the scheme-dependent additive constant~\cite{Humpert}.

In the off-shell regime, the evolution equation for the scattering amplitude\\
$\mathcal{M}_4^{\rm off-shell} = 1 + a M^{(1)}_{ \rm off-shell} + O(a^2)$ takes
the form~\cite{Korchemsky2}
\be\label{cusponeoff}
\left(\frac{\pa}{\pa \ln m^2}  \right)^2 \ln\mathcal{M}_4^{\rm off-shell} =
-2\Gamma_{\rm cusp}(a) + O(m)\,,
\ee
giving again  $\Gamma_{\rm cusp} = 2a + O(a^2)$. Notice the characteristic
difference of a factor of 2 in (\ref{cusponeoff}) compared to the on-shell
expression (\ref{defcusp}). Its origin can be understood as
follows~\cite{Korchemsky2}. The amplitude $M^{(1)}_{ \rm off-shell}$ is given by
the one-loop scalar box integral in which the loop momentum $k^\mu$ is integrated
over the whole phase space.   The divergent contribution  to $M^{(1)}_{ \rm
off-shell}$, Eq.~(\ref{divepartoff}), comes from two regions in the phase space:
the so-called soft region, $k^\mu = O(m)$, and the infrared (or `ultra-soft'
\cite{Kuhn99}) region, $k^\mu = O(m^2/Q)$, with the hard scale $Q\sim
\sqrt{|s|}, \sqrt{|t|}$.
Each region produces a double-logarithmic contribution $\sim (\ln m^2)^2$ which
translates into $(-\Gamma_{\rm cusp}(a))$ in the right-hand side of
(\ref{cusponeoff}). This explains the factor 2 in (\ref{cusponeoff}). In the
on-shell regime, one finds that the infrared region does not exist while the soft
region provides the same $(-\Gamma_{\rm cusp}(a))$ contribution to the right-hand
side of the evolution equation (\ref{defcusp}).

\subsection{Generalization to all orders}

In this subsection we give a review of some generic properties of gluon
amplitudes in gauge theory and some special properties of the four-gluon
amplitudes in $\cN=4$ SYM. The first property we will need in the following
discussion is the factorization of the infrared singular amplitude, both on and
off shell, into Sudakov form factors $M_{gg\rightarrow 1}$ for a colorless boson
decaying into two gluons. Further, the singularities of the latter are known to
exponentiate~\cite{Sudakov54,Gatheral}, thus determining the structure of the
infrared singularities of the four-gluon amplitude up to a few constants.
However, this leaves the finite part of the four-gluon amplitude arbitrary. At
the end of this subsection we show that the additional assumption of dual
conformal symmetry in the off-shell regime fixes the finite part up to an
additive constant.

\subsubsection{On-shell regime}

In the on-shell regime ($p^2_i=0$, $D=4-2\ep$) and in the planar limit the  $n$-gluon amplitude takes the
factorized form
\begin{equation}\label{ncake}
    \mathcal{M}_n = \mathcal{F}_n \prod_{i=1}^n \left[M_{gg \rightarrow 1}\left(\frac{\mu_{\IR}^2}{-s_{i,i+1}},a,\ep\right)\right]^{1/2}\ .
\end{equation}
Here the factor $\mathcal{F}_n$ is finite and the kinematic variables are
$s_{i,i+1} = (p_i + p_{i+1})^2$.

Restricting (\ref{ncake}) to the case of interest $n=4$, we can write the factorized amplitude in the following form:
\begin{equation}\label{ncake'}
\mathcal{M}_4^{\rm on-shell}  =  \mathcal{F}_4^{\rm on-shell}\ M^{\rm
on-shell}_{gg \rightarrow 1}\left(\frac{\mu_{\IR}^2}{-s},\ep\right)\ M^{\rm
on-shell}_{gg \rightarrow 1}\left(\frac{\mu_{\IR}^2}{-t},\ep\right)\ .
\end{equation}

The form factor $M^{\rm on-shell}_{gg \rightarrow 1}$ satisfies an evolution
equation~\cite{on-shell}. In a finite theory, where the coupling does not run,
the solution of this evolution equation has a particularly simple exponential
form:
\begin{equation}\label{ffon}
\ln M^{\rm on-shell}_{gg\rightarrow 1}\left(\frac{\mu_{\IR}^2}{-s}\right) =
-\frac{1}{2} \sum_{l=1}^{\infty} a^l \left(
\frac{\mu_{\IR}^2}{-s}\right)^{l\ep}\left[\frac{\Gamma^{(l)}_{\rm
cusp}}{(l\ep)^2} + \frac{G^{(l)}}{l\ep} + A^{(l)}\right] + O(\ep)\ .
\end{equation}
Here $\Gamma^{(l)}_{\rm cusp}$ are the coefficients in the perturbative expansion
of the cusp anomalous dimension $\Gamma_{\rm cusp}(a)$. The constants $G^{(l)}$
and $A^{(l)}$ are {regularization} scheme dependent.

Replacing $M^{\rm on-shell}_{gg \rightarrow 1}$ in (\ref{ncake'}) by its
expression (\ref{ffon}), we find the following splitting of the log of the
four-gluon amplitude into a divergent and a finite parts:
\begin{equation}\label{splitongen}
\ln \mathcal{M}^{\rm on-shell}_4 = \ln M^{\rm on-shell}_{gg\rightarrow
1}\left(\frac{\mu_{\IR}^2}{-s},\ep \right) + \ln M^{\rm on-shell}_{gg\rightarrow
1}\left(\frac{\mu_{\IR}^2}{-t},\ep \right) + \ln {\cal F}^{\rm
on-shell}_4\left(\frac{s}{t}\right) + O(\ep)\ .
\end{equation}
{This relation generalizes the one-loop result (\ref{M1-sum}) to higher
loops.  A unique feature of this particular splitting of $\ln \mathcal{M}^{\rm
on-shell}_4$ is that } the $\mu_{\IR}^2$ dependence is restricted to the
divergent part of the amplitude (coming from the form factors). From (\ref{splitongen}) one can
extract the cusp anomalous dimension applying (\ref{defcusp}). At the same time,
the finite part, being independent of the IR scale $\mu_{\IR}^2$, is a function
of the remaining dimensionless variable $s/t$ only.

One of the central results of Ref. \cite{bds05} is the conjecture that in the case of $\cN=4$ SYM the finite part of the four-gluon amplitude, as defined by the splitting (\ref{splitongen}), has a very simple all-order form:
\begin{equation}\label{central}
\ln {\cal F}^{\rm on-shell}_4 = \frac{\Gamma_{\rm cusp}(a)}{4} \ln^2 \frac{s}{t}
+ \mbox{const}\ .
\end{equation}
One may say that the one-loop finite part (\ref{fponsh}) exponentiates to all orders.
This conjecture has been verified in \cite{bds05} up to three loops. The same form of the finite part has been found in \cite{am07} at strong coupling. In what follows we give an argument in favor of this conjecture, based on the assumption of conformal invariance of the amplitude in the off-shell regime.

\subsubsection{Off-shell regime}

In the off-shell regime ($p^2_i=-m^2$, $D=4$) the four-gluon amplitude still
factorizes as indicated in (\ref{ncake'}), but now the Sudakov form
factor~\cite{Sudakov54,Smilga79} has the following infrared divergent structure
\cite{Korchemsky2}:
\begin{equation}\label{0}
\ln M^{\rm off-shell}_{gg\rightarrow 1}\left(\frac{m^2}{-s}\right) = -\frac{1}{2}
\Gamma_{\rm cusp}(a)\ \ln^2 \left(\frac{m^2}{-s}\right) + G^{\rm off-shell}(a) \ln
\left(\frac{m^2}{-s}\right) + {\rm const} + O(m)\ .
\end{equation}
The main difference from the on-shell regime is that $\Gamma_{\rm cusp}$ appears
with a factor of 2 off shell (compare the double-log derivatives of (\ref{ffon})
and (\ref{0}) with respect to the IR scale). The reason for this is the same as
for the one-loop off-shell amplitude (\ref{cusponeoff}) -- the Sudakov form
factor receives an additional contribution from the infrared region $k^\mu =
O(m^2/(\sqrt{-s}))$.

The splitting of the four-gluon amplitude takes the form (cf. (\ref{splitongen}))
\begin{equation}\label{01}
\ln \mathcal{M}_4^{\rm off-shell} = \ln M^{\rm off-shell}_{gg\rightarrow
1}\left(\frac{m^2}{-s}\right)  + \ln M^{\rm off-shell}_{gg \rightarrow 1}
\left(\frac{m^2}{-t} \right)  + \ln {\cal F}_4^{\rm
off-shell}\left(\frac{s}{t}\right) + O(m)\ .
\end{equation}
{This relation generalizes the one-loop result (\ref{splitoff}) to higher
loops.} Substituting (\ref{0}) into (\ref{01}), we obtain
\begin{eqnarray}
  \ln \mathcal{M}_4^{\rm off-shell} &=& -\frac{1}{4} \Gamma_{\rm cusp}(a) \ln^2 \frac{m^4}{st} +G^{\rm off-shell}(a) \ln \frac{m^4}{st} \nn\\
   && - \frac{1}{4} \Gamma_{\rm cusp}(a) \ln^2\frac{s}{t}  + \ln {\cal F}_4^{\rm off-shell}\left(\frac{s}{t}\right) + {\rm const} + O(m)\ . \label{inspect}
\end{eqnarray}

\subsubsection{Off-shell conformal symmetry and the form of the finite part}

As we have seen earlier, there is substantial evidence from perturbation theory
that the planar four-gluon amplitude in the on-shell regime possesses
a hidden
dual conformal structure. We may take this as an indication that the
full off-shell amplitude discussed above will exhibit the same
symmetry. As discussed
in subsection \ref{dual} this is not obvious because the full off-shell
amplitude may contain additional integrals which vanish in the
on-shell regime.
Nevertheless, let us adopt this conformal symmetry as an assumption.
Then, up to terms which vanish as $m \to 0$, the amplitude $\mathcal{M}_4^{\rm off-shell}$
can only depend on the conformal cross-ratios $u$ and $v$ which, for the special
choice $p^2_i=-m^2$, are given by (\ref{crossoffsh}). Thus, the only conformally
invariant variable is ${m^4}/{st}$, while the ratio $s/t$ is not conformal. We
are then lead to the conclusion that the term in  (\ref{inspect}), involving
$\ln^2 ({s}/{t})$ and originating from the form factors, {must cancel against a similar
term coming from ${\cal F}_4^{\rm off-shell}$}. Moreover, any further dependence
on $s/t$ in ${\cal F}_4^{\rm off-shell}$ is ruled out by the assumption of dual
conformal invariance. In this way, we
arrive at
\begin{equation}
\ln {\cal F}_4^{\rm off-shell}\left(\frac{s}{t}\right) = \frac{\Gamma_{\rm
cusp}(a)}{4}  \ln^2 \frac{s}{t} + {\rm const}\ ,
\end{equation}
which is precisely the form (\ref{central}) observed in \cite{bds05} in
perturbation theory and in \cite{am07} at strong coupling. We can interpret this
as a manifestation of the dual conformal structure of four-gluon
amplitudes.

\pagebreak

\section{Light-like Wilson loops}

In this section we discuss the relation between the four-gluon scattering amplitude
and the expectation value of a Wilson loop,
\begin{equation}\label{W}
    W_C = \frac1{N}\vev{0|{\rm Tr}\, P \exp\lr{ig\int_C dx^\mu A_\mu(x)} |0}\,.
\end{equation}
Here $ A_\mu(x)=A_\mu^a(x) t^a$ is a gauge field and $t^a$ are the $SU(N)$
generators in the fundamental representation, $C$ is a closed contour in
Minkowski space and $P$ stands for path ordering of the $SU(N)$ indices. To the lowest order in the coupling it is given by
\be\label{W-one-loop}
W_C = 1 + \frac12(ig)^2C_F \int_C dx^\mu \int_C dy^\nu \, G_{\mu\nu}(x-y) +
O(g^4)
\ee
where $C_F=(N^2-1)/(2N)$ is the Casimir of the fundamental representation of $SU(N)$ and
$G_{\mu\nu}(x-y)$ is the free gluon propagator in a particular gauge. Since the
Wilson loop (\ref{W}) is a gauge invariant functional of the integration contour
$C$, we can choose the gauge for convenience.

Following \cite{am07}, we choose the integration contour in (\ref{W}) to consist
of four light-like segments joining the points $x_i^\mu$ (with $i=1,2,3,4$) such
that $x_i-x_{i+1}=p_i$ coincide with the external on-shell momenta of the
four-gluon scattering amplitude (recall the relation (\ref{chva})). We would like
to stress that this Wilson loop has the following unusual feature -- the
integration contour $C$ is defined by the external momenta of the four-gluon
scattering amplitude, but the gluon propagator in (\ref{W-one-loop}) is the one
in configuration space, not in momentum space. Similar Wilson loops have already
been discussed in the past in the context of the infrared asymptotics of the QCD
scattering amplitudes and their properties were studied in
Refs.~\cite{KR86,IKR85,KK92,K88}.

Due to Lorentz invariance, the Wilson loop $W_C$ is a function of the scalar products
$(x_i\cdot x_j)$. The conformal symmetry of $\mathcal{N}=4$ SYM imposes additional
constraints on the possible form of this function. If the Wilson loop
$W(C)$ were
well defined in four dimensions, one would deduce that it changes
under the $SO(2,4)$ conformal transformations (translations,
rotations, dilatations and special conformal transformations) as
$W_C = W_{C'}$ where the contour $C'$ is the image of $C$ under these
transformations. In this way, one would conclude that $W(C)$ could only depend on the
two conformal invariant cross-ratios $u$ and $v$ defined in (\ref{cross}).
However, for the Wilson loop under consideration the conformal symmetry becomes
anomalous due to the presence of cusps on the integration contour $C$. These cusps
lead to UV divergences in $W_C$ which need to be regularized.

\subsection{Cusp anomaly}

To illustrate the cusp anomaly, let us evaluate the double contour integral in
(\ref{W-one-loop}). In what follows we shall employ dimensional regularization
$D=4-2\vep$ (with $\vep>0$; notice the difference with the IR regulator $\ep<0$)
and use the notation of \cite{KK92}. By virtue of gauge invariance, we can choose
the Feynman gauge in which the gluon propagator has the form
\be
G_{\mu\nu}(x) =  -g_{\mu\nu}\frac{\Gamma(1-\vep)}{4\pi^{2}} (-
x^2+i0)^{-1+\vep}\lr{\pi \tilde\mu^2}^{\vep}
\ee
and perform the calculation directly in configuration space. For a reason which
will become clear in a moment, we introduced the notation $\tilde\mu^2$ for dimensionful
parameter to distinguish it from the similar parameter $\mu^2$ that
we used to regularize IR divergences in Section~3. Splitting the integration
contour into four segments $C=C_1\cup C_2\cup C_3\cup C_4$ and introducing the
parametrization $C_i=\{x^\mu(\tau_i) = x^\mu_i - \tau_i p_i^\mu|\,0\le \tau_i
\le 1\}$, we find from (\ref{W-one-loop})
\be\label{W=I}
\ln W_C=-\frac{g^2 C_F}{4\pi^2} \sum_{1\le j\le k \le 4} I_{jk} +O(g^4)
\ee
where
\be\label{I-integral}
I_{jk} =-\int_0^1 d\tau_j \int_0^1 d\tau_k\frac{(p_j\cdot
p_k)\Gamma(1-\vep)\lr{\pi \tilde\mu^2}^{\vep}}{[-(x_j-x_k - \tau_j p_j+\tau_k
p_k)^2+i0]^{1-\vep}}
\ee
with $x_i-x_{i+1}=p_i$.
\begin{figure}[h]
\psfrag{x1}[cc][cc]{$x_1$}
\psfrag{x2}[cc][cc]{$x_2$}
\psfrag{x3}[cc][cc]{$x_3$}
\psfrag{x4}[cc][cc]{$x_4$}
\psfrag{a}[cc][cc]{(a)}
\psfrag{b}[cc][cc]{(b)}
\psfrag{c}[cc][cc]{(c)}
\centerline{{\epsfysize4cm \epsfbox{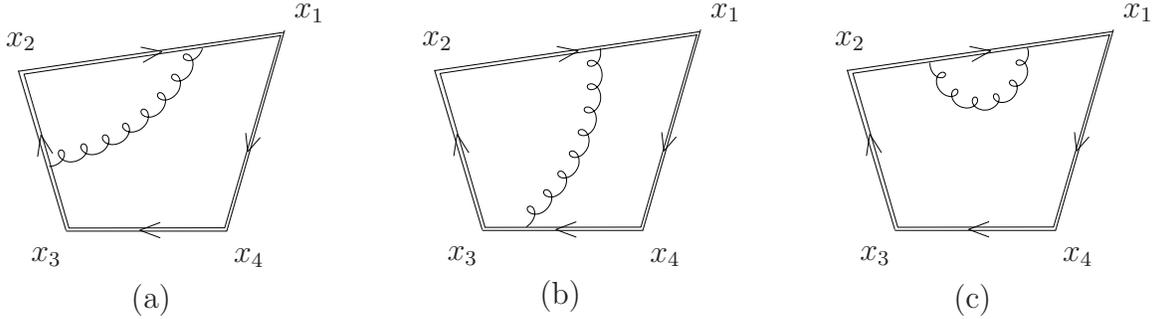}}} \caption[]{\small The Feynman diagram
representation of the integrals (\ref{I-integral}): (a) $I_{12}$, (b) $I_{13}$
and (c) $I_{11}$. The double line depicts the integration contour $C$ and the
wiggly line the gluon propagator.}
\label{Fig-WL}
\end{figure}
It is convenient to represent these integrals by the Feynman diagrams shown
in Fig.~\ref{Fig-WL}.

Then, to one-loop accuracy we encounter three types of integrals depicted in
Fig.~\ref{Fig-WL}(a), (b) and (c). We begin with the last one and take into
account the on-shell condition $p_j^2=0$ to find $I_{jj}\sim p_j^2=0$. Let us now
examine the integral $I_{12}$ corresponding to Fig.~\ref{Fig-WL}(a)
\be
I_{12} = -\int_0^1 d\tau_1 \int_0^1 d\tau_2\frac{(p_1\cdot
p_2)\Gamma(1-\vep)\lr{\pi \tilde\mu^2}^{\vep}}{[-2(p_1\cdot p_2)(1 -
\tau_1)\tau_2]^{1-\vep}}\ ,
\ee
so that the integration over $\tau_1$ and $\tau_2$ yields
\be\label{I12}
I_{12} = \lr{\pi\tilde\mu^2(- s)}^\vep\frac{\Gamma(1-\vep)}{2\epsilon_\UV^2}\ .
\ee
Here the double pole in $\vep$ comes from integration in the vicinity of the cusp
located at  point $x_2$ and has a clear ultraviolet origin.\footnote{By `UV' here
we mean small distances in the dual `configuration' space of the $x_i$.} It is
easy to see from (\ref{I-integral}) that $I_{34}=I_{12}$ whereas the integrals
$I_{23} = I_{14}$ can be obtained from $I_{12}$ by substituting $s\mapsto t$ with
$s= (p_1+p_2)^2$ and $t=(p_2+p_3)^2$ being the usual Mandelstam variables in the
momentum space of the four-gluon amplitude. The remaining integrals $I_{13} =
I_{24}$ are computed in a similar manner. We first verify that the integral
$I_{13}$ depicted in Fig.~\ref{Fig-WL}(b) remains finite for $\vep\to 0$ and,
therefore, can be evaluated in $D=4$. In this way, we get
\be
I_{13}=  \int_0^1 d\tau_1 \int_0^1 d\tau_3\frac{(p_1\cdot p_3) }{[p_1(1 -
\tau_1)+p_2+p_3\tau_3]^2} = -\frac12\int_0^1 d\tau_1 \int_0^1
d\tau_3\frac{s+t}{s\bar\tau_1+t\tau_3+st\bar\tau_1\tau_3 }
\ee
with $\bar\tau_1=1-\tau_1$. The integration can be easily performed and yields
\be\label{I13}
I_{13}= - \frac14  \left[\ln^2 ({s}/{t})+\pi^2\right] + O(\vep)\,.
\ee
We can then replace the integrals $I_{ik}$ in (\ref{W=I}) by their expressions
(\ref{I12}) and (\ref{I13}) and evaluate the one-loop correction to the
light-like Wilson loop. Notice that in Eq.~(\ref{I12}) the scale $\tilde\mu^2$
has the `wrong' dimension of [mass]${}^{-2}$ as compared to the conventional
dimensional regularization scale. This is due to the fact that the
contour $C$ is defined in terms of the dual coordinates $x_i$, which in turn are
related to the momenta $p_i$ through (\ref{chva}). It is then convenient to
introduce the new scale
\be
\mu_{\UV}^2 = \lr{\widetilde\mu^2\pi{\rm e}^\gamma}^{-1}
\ee
and to rewrite the one-loop expression for the Wilson loop (\ref{W=I}) in the
multi-color limit as follows
\be
\ln W_C = a\left\{-\frac{1}{\epsilon^2_\UV} \left[ \left(
\frac{\mu_{\UV}^2}{-s}\right)^{-\vep} +
\left(\frac{\mu_{\UV}^2}{-t}\right)^{-\vep} \right]+ \frac12\ln^2\frac{s}{t}
+2\zeta_2+ O(\vep) \right\} + O(a^2)\,.
\ee
Comparing this relation with the one-loop expression for the four-gluon
scattering amplitude, Eqs.~(\ref{M1-sum}) -- (\ref{fponsh}), we observe that,
firstly, the divergent parts of the two expressions coincide provided that we
formally identify the UV cutoff for the Wilson loop with the IR cutoff for the
scattering amplitude, $\mu_{\UV}^2 = \mu_{\IR}^2$, and the IR regulator  $\ep$
with the UV one $\vep$  (remembering, however, that $\ep$ and $\vep$ have
different signs), $\ep=-\vep$. Secondly, the finite part of the Wilson loop
contains the same $\ln^2(s/t)$ term as the scattering amplitude (\ref{fponsh}),
while the additive constants are different.

We can interpret this result as an indication that the duality between four-gluon
scattering amplitudes and light-like Wilson loops proposed at strong coupling by
Alday and Maldacena \cite{am07} also exists at weak coupling. However, the
explicit form of the duality transformation which relates the two objects in
$\mathcal{N}=4$ SYM remains to be found.

\subsection{Evolution equations}

A natural question to ask is whether the same duality between the four-gluon
scattering amplitude and the light-like Wilson loop will survive at higher loops
in the planar limit. For the divergent part of the two quantities the answer can
be found by making use of the evolution equations. For the on-shell four-gluon
scattering amplitude, one finds from (\ref{splitongen}) that its dependence on
the IR scale $\mu_{\IR}$ is given by
\be\label{M-evol}
\frac{\pa}{\pa\ln \mu_{\IR}^2} \ln \mathcal{M}_4= -\frac12 \Gamma_{\rm cusp}
(a)\ln\lr{\frac{\mu_{\IR}^4}{st}} - G(a) -
\frac1{\ep}\int_0^a\frac{da'}{a'}\Gamma_{\rm cusp} (a') + O(\ep)\ ,
\ee
where $G(a) = \sum_l a^l \, G^{(l)}$ is the so-called collinear anomalous
dimension. It has been calculated in \cite{bds05} to three-loop accuracy:
\be\label{G-3}
G(a) = - \zeta_3 a^2 +  \lr{4\zeta_5+\frac{10}3\zeta_2\zeta_3 } a^3+
O(a^4)\,.
\ee
The three-loop expression for the cusp anomalous dimension reads \cite{bds05}
\be\label{cusp-3}
\Gamma_{\rm cusp}(a) = 2 a -2 \zeta_2 a^2 + 11 \zeta_4 a^3 + O(a^4)\,.
\ee

For the light-like Wilson loop {under consideration}, a similar relation
follows from its renormalization properties~\cite{KK92}:
\be\label{Ward}
\frac{\pa}{\pa\ln\mu_{\UV}^2} \ln W_C= -\frac12 \Gamma_{\rm cusp}
(a)\ln\lr{\frac{\mu_{\UV}^4}{st}} - \Gamma(a) +
\frac1{\vep}\int_0^a\frac{da'}{a'}\Gamma_{\rm cusp} (a') + O(\vep)
\ee
with the anomalous dimension $\Gamma(a)=0\cdot a + O(a^2)$. The relation
(\ref{Ward}) can also be interpreted as the dilatation Ward identity for the
Wilson loop in $\mathcal{N}=4$ SYM. Comparing relations (\ref{M-evol}) and
(\ref{Ward}), we conclude that the IR divergent part of $\ln \mathcal{M}_4$
matches the UV divergent part of the dual light-like Wilson loop to all orders
provided that the corresponding scales are related to each other as follows:
\be\label{scales}
\ln \frac{\mu_{\UV}^2}{\mu_{\IR}^2} =  \frac{G(a)-\Gamma(a)}{\Gamma_{\rm cusp}
(a)}\ .
\ee
Since the anomalous dimensions $G(a)$ and $\Gamma(a)$ receive perturbative
corrections starting from two loops only, the right-hand side of this relation is
given by a series in the coupling constant $a$.

To evaluate the normalization factor entering the scale-setting relation
(\ref{scales}), one has to determine the Wilson loop anomalous dimension
$\Gamma(a)$ to two loops and supplement it with the known results for $G(a)$ and
${\Gamma_{\rm cusp} (a)}$. Since $\Gamma(a)$ does not depend on the Mandelstam
variables $s,\,t$, we may simplify the analysis by choosing $t=-s$, or
equivalently $p_2=-p_3$ and $p_1=-p_4$. It is easy to see that the resulting
integration contour $C$ takes the form of a rhombus with its parallel sides along
two different light-cone directions. Such a Wilson loop has been studied in QCD
in the context of the gluon Regge trajectory~\cite{KK96} (see the next section)
and the same two-loop expression for $W_C$ has been found in two different
gauges~\cite{KK92} (Feynman and light-like axial gauge). It is straightforward to
generalize the two-loop QCD result to $\mathcal{N}=4$ SYM~\cite{BGK03}. To this
end one has to add to $W_C$ the contribution of $n_s=6 N$ scalars, $n_f=4 N$
gauginos (to two loops, they only enter through the one-loop correction to the
gluon polarization operator) and convert the result from the dimensional
regularization scheme (DREG) to the dimensional reduction scheme (DRED). In this
way we have found that for $s=-t$ the two-loop Wilson loop $W_C$ satisfies the
evolution equation (\ref{Ward}) with the following value of the two-loop
anomalous dimension
\be
\Gamma(a) = - 7\zeta_3 a^2 + O(a^3)\,.
\ee
Substituting this relation into (\ref{scales}) and taking into account
(\ref{G-3}) and (\ref{cusp-3}), we finally find
\be\label{scales-1}
\ln \frac{\mu_{\UV}^2}{\mu_{\IR}^2} = 3\zeta_3 a + O(a^2)\,.
\ee
Under such an identification of the scales, the light-like Wilson loop $\ln W_C$ matches
the on-shell four-gluon scattering amplitude $\ln \mathcal{M}_4(s,t=-s)$ to two
loops up to finite, $s-$independent constant terms.

Obviously, the calculation of the Wilson loop for $s=-t$ does not allow us
to test the form of the finite part. It would be interesting to carry out a full
two-loop calculation of the Wilson loop to see if the duality gluon
amplitudes/Wilson loops also applies to the functional dependence on $s,t$, as it
did at one loop.

\subsection{Conformal invariance of the dual Wilson loops}

As was already mentioned, the cusp anomaly breaks the $SO(2,4)$ conformal
symmetry of the four-dimensional Wilson loops. If this anomaly was not present,
the Wilson loop would be a function of the conformally invariant cross-ratios
(\ref{cross}) only. Notice that the cusp anomaly originates from the integration
in the vicinity of the cusps and, as a consequence, its contribution depends on
the corresponding cusp angle or, equivalently, on the scalar products $2(p_i\cdot
p_{i+1})$. In other words, each cusp produces an additive contribution to $\ln
W_C$ depending either on $s$, or on $t$ but not on both variables simultaneously.
This suggests that the `crossed' terms $\sim \ln (-s) \ln(-t)$ in the
perturbative expansion of $\ln W_C$ will not be affected by the cusp anomaly and,
therefore, their form is still subject to the conformal symmetry constraints.
Indeed, we already observed that the finite $\ln^2(s/t)$ contribution to $\ln
W_C$ matches the similar contribution to $\ln \mathcal{M}_4$, which in turn
follows from the conjectured off-shell conformal symmetry of the scattering
amplitude. In the dual, Wilson loop description this symmetry is just the
conformal symmetry of the four-dimensional Wilson loops.

Above we have demonstrated that the IR divergences of the on-shell scattering
amplitude are dual to the UV (cusp) divergences of the light-like Wilson loop.
One may wonder whether the same correspondence would survive if one assigned
off-shellness (virtuality) to the momenta, $p_i^2=-m^2$. In this case, the
off-shell scattering amplitude is well defined in four dimensions while the cusp
anomaly is still present in the Wilson loop. It produces a divergent contribution
to $W_C$ which satisfies the following evolution equation
\be
\frac{\pa}{\pa\ln\mu_{\UV}^2} \ln W_C^{\rm off-shell}= -\frac12 \Gamma_{\rm cusp}
(a)\ln\lr{\frac{m^4}{st}} + O(\vep)\,.
\ee
This implies that the (finite) off-shell scattering amplitude $\mathcal{M}_4^{\rm
off-shell}$ cannot be identified with the (divergent) Wilson loop $W_C^{\rm
off-shell}$ evaluated along the same contour $C$ as before with the only
difference being that $p_i^2=-m^2$, or equivalently, that the segments
run along
space-like directions. {This indicates that the precise nature of the duality
between off-shell gluon amplitudes and Wilson loops needs to be investigated in
more depth.}

\section{Gluon Regge trajectory}

In this section we examine the asymptotic behaviour of the on-shell  four-gluon
scattering amplitude (\ref{splitongen}) in the Regge limit
\be\label{Regge}
s >0\,, \quad t<0 \,, \quad s\gg -t\,.
\ee
It is expected that the scattering amplitude in this limit is given by the sum
over Regge trajectories, each producing a power-like contribution
$\mathcal{M}_4(s,t) \sim s^{\omega(-t)}$. In the planar limit, only the
trajectory with the quantum numbers of a gluon gives a dominant contribution, so
that the sum over the Regge trajectories is reduced to a single contribution from
the gluon Regge trajectory  \cite{Schnitzer,Kuraev76},
\be\label{Regge-ansatz}
\mathcal{M}_4(s,t) =  [c(-t)]^2 \lr{\frac{s}{-t}}^{\omega_R(-t)} +
\mbox{[subleading terms in $|t|/s$]} \ ,
\ee
where $\omega_R(-t)$ is the Regge trajectory and $c(-t)$ is the gluon impact
factor. 
It is believed that the relation (\ref{Regge-ansatz}) holds in generic
gauge theories ranging from QCD to $\mathcal{N}=4$ SYM. Indeed, the gluon
reggeization was first discovered in QCD \cite{Kuraev76} and the gluon Regge
trajectory is presently known to two-loop accuracy~\cite{Fadin96}. Moreover, it
has been shown in \cite{KK96} that the all-loop gluon Regge trajectory takes the
following form in QCD
\be\label{QCD-regge}
\omega^{\rm (QCD)}_R(-t) =   \frac12 \int_{(-t)}^{\mu_{\IR}^2}
\frac{dk_\perp^2}{k_\perp^2} \Gamma_{\rm cusp}(a(k_\perp^2)) + \Gamma_{R}(a(-t))
+ \mbox{[poles in $1/\ep$]}\ ,
\ee
where the two terms on the right-hand side define finite (as $\ep\to 0$)
contributions. .
Also, since the QCD beta-function is different from zero, the coupling constant
depends on the normalization scale as indicated in (\ref{QCD-regge}). The
anomalous dimension $\Gamma_{R}(a)$ is given to two loops by
\be\label{Gamma-R}
\Gamma_{R}(a)= 0\cdot a + a^2 \left[\frac{101}{27}-\frac12\zeta_3-\frac{14}{27}
\frac{n_f}{N} \right]+O(a^3)
\ee
where $n_f$ is the number of quark flavors.

We would like to stress that for generic values of the Mandelstam variables the
four-gluon QCD scattering amplitude has a rather complicated form~\cite{Glover01}
different from (\ref{Regge-ansatz}). It is only in the Regge limit that one
recovers (\ref{Regge-ansatz}) as describing the leading asymptotic behaviour of
the four-gluon QCD scattering amplitude in the planar
approximation~\cite{DelDuca01}. One would expect that a similar simplification
should also take place for the four-gluon planar amplitude in $\mathcal{N}=4$ SYM
in the Regge limit. As we will see in a moment, this amplitude has the remarkable
property of being {\it Regge exact}, i.e. the contribution of the gluon Regge
trajectory to the amplitude (\ref{Regge-ansatz}) coincides with the exact
expression for $\mathcal{M}_4(s,t)$ with $s$ and $t$ arbitrary.

If the relation (\ref{Regge-ansatz}) is exact, i.e. if the subleading terms
in the right-hand side of (\ref{Regge-ansatz}) are absent, then the scattering
amplitude has to satisfy the evolution equation
\be
\frac{\pa}{\pa  \ln s}\ln \mathcal{M}_4(s,t) = \omega_R(-t)\ .
\ee
Notice that the right-hand side of this relation should be $s-$independent up to
terms vanishing as $\ep \to 0$. Let us verify this relation using the factorized
expression (\ref{splitongen}) for $\ln \mathcal{M}_4$:
\be
\omega_R(-t)=\frac{\pa}{\pa  \ln s}\left[ \ln M^{gg\rightarrow
1}\left(\frac{\mu_\IR^2}{-s},\ep \right)  + \ln {\cal F}_{\rm
on-shell}\left(\frac{s}{t}\right)   \right]\ .
\ee
Taking into account (\ref{ffon}) and (\ref{central}) we find that, in agreement
with our expectations, the $s-$dependence disappears in the sum of the two terms
and we obtain the following gluon trajectory:
\be\label{N4-regge}
\omega_R(-t) =  \frac12 \Gamma_{\rm cusp} (a)\ln {\frac{\mu_{\IR}^2}{(-t)}}
+\frac12 G(a) + \frac1{2\ep}\int_0^a\frac{da'}{a'}\Gamma_{\rm cusp} (a') +
O(\ep)\ .
\ee
Together with (\ref{G-3}) and (\ref{cusp-3}), this relation defines the gluon
Regge trajectory in $\mathcal{N}=4$ SYM to three loops.

{Let us compare the expressions for the gluon trajectory in QCD and in
$\mathcal{N}=4$ SYM, Eqs.~(\ref{QCD-regge}) and (\ref{N4-regge}), respectively.
It is easy to see that (\ref{N4-regge}) can be obtained from (\ref{QCD-regge}) if
we neglect the running of the coupling constant (recall that the beta function
vanishes in $\mathcal{N}=4$ SYM to all loops) and identify  the anomalous
dimension $\frac12G(a)$ with $\Gamma_{R}(a)$ in $\mathcal{N}=4$ SYM. Moreover,
comparing the two-loop expression (\ref{Gamma-R}) for $\Gamma_{R}(a)$ with
$\frac12G(a) = - \frac12\zeta_3 a^2 + O(a^3)$ (see (\ref{G-3})), we observe that
the latter can be obtained from the former by retaining the terms of maximal
transcendentality only.

The analysis performed in Section~4 suggests that in $\mathcal{N}=4$ SYM at weak
coupling the four-gluon planar amplitude $\ln \mathcal{M}_4(s,t)$ matches, for
arbitrary $s$ and $t$, the expectation value of the light-like Wilson loop $\ln
W_C$ up to an additive constant term. Then, going to the Regge limit
(\ref{Regge}) and making use of the relation (\ref{Regge-ansatz}), we can
identify the gluon Regge trajectory in terms of the dual Wilson loop. The
relation between these two seemingly different objects was first
observed in QCD in Ref.~\cite{KK96}.
\footnote{We would like to stress that, in distinction with $\mathcal{N}=4$ SYM,
the four-gluon planar amplitude in QCD is not dual to light-like Wilson loop for
arbitrary $s$ and $t$. The relation between the two quantities emerges in the
Regge limit only.} The Wilson loop (\ref{W}) is defined in QCD in the same way as
in $\mathcal{N}=4$ SYM. At weak coupling, the one-loop expressions for $W_C$ are
the same in the two theories but they differ from each other starting from two
loops. It was observed in \cite{KK96} that the two-loop QCD expression for the
Wilson loop $W_C$, with $C$ being a light-like rectangular loop, takes the Regge-like
form (\ref{Regge-ansatz}). Moreover, the resulting two-loop expression for the
exponent of $s$ can be brought to the same form as the two-loop gluon Regge
trajectory (\ref{QCD-regge}). This is achieved by replacing the IR cutoff $\mu_{\IR}^2$ by the UV cutoff
$\mu_{\UV}^2$, and the two-loop anomalous dimension $\Gamma_{R}(a)$
in the right-hand side of (\ref{QCD-regge}) by
\be
\Gamma_{R}^{\rm (dual)}(a) =   0\cdot a + a^2
\left[\frac{101}{27}-\frac72\zeta_3-\frac{14}{27} \frac{n_f}{N_c}
\right]+\mathcal{O}(a^3)\,.
\ee
Then we express $\mu_{\UV}^2$ in terms of $\mu_{\IR}^2$ with the help of
(\ref{scales}) and take into account the relation $\Gamma_{R}(a)-\Gamma_{R}^{\rm
(dual)}(a)=3\zeta_3\, a^2 + O(a^3)$ to verify that the Regge trajectory obtained
from the Wilson loop calculation coincides with the two-loop expression for the
gluon Regge trajectory in QCD (\ref{QCD-regge}).

Needless to say, QCD is very different from $\mathcal{N}=4$ SYM and its dual
string description is not known yet. Nevertheless, the very fact that the
two-loop gluon Regge trajectory in QCD admits a dual description in terms of a
light-like Wilson loop provides yet another indication that QCD possesses some
hidden (integrable) structure~\cite{Belitsky04}. }

\section{Summary and discussion}\label{disc}

In this paper we have presented further evidence for a dual conformal symmetry in
the planar four-gluon amplitude in $\cN=4$ SYM. We have shown that all the momentum loop
integrals appearing in the perturbative calculations up to five loops are dual to
true conformal integrals, well defined off shell. Assuming that the complete
off-shell amplitude has this property, we have derived the special form of the
finite remainder previously found perturbatively and reproduced at strong
coupling by AdS/CFT. We have also shown that the same finite term appears in a
weak coupling calculation of a Wilson loop whose contour consists of four
light-like segments associated with the gluon momenta. {We have demonstrated
that the dual conformal symmetry leads to dramatic simplification of the planar
four-gluon amplitude in the high-energy (Regge) limit. Namely, due to the special
form of the finite remainder, the contribution of the gluon Regge trajectory to
the amplitude coincides with its exact expression evaluated for arbitrary values
of the Mandelstam variables.}

Several questions remain open. First of all, we need to investigate the
four-gluon amplitude in the off-shell regime beyond one loop. One should not take
it for granted that going off shell simply consists in changing the regulator in
the set of loop integrals contributing to the on-shell amplitude. It is quite
likely that new integrals will appear off shell, such that they vanish for
$p^2_i=0$ in dimensional regularization but contribute when $p^2_i\to 0$ in
$D=4$. There are good reasons to believe that this will indeed be the case
\cite{London,Dixon}. The crucial question will then be whether the dual conformal
symmetry concerns these additional contributions as well. A two-loop calculation
of the off-shell amplitude, which may help clarify this point, is currently under
way.

Another question is if the dual conformal symmetry is specific to four-gluon
amplitudes or also applies to multi-gluon ones. The iteration conjecture of
\cite{bds05} predicts the exponentiation of the one-loop finite part
independently of the number of external legs. This has so far been confirmed by
an explicit two-loop five-gluon calculation in \cite{Cachazo:2006tj,5point}. One has to analyze the
integrals appearing there to see if they possess similar conformal properties. It
should be pointed out that the number of conformally invariant variables
(cross-ratios) rapidly grows with the number of points, therefore multiple-point
dual conformal symmetry, if present, may be less restrictive than at four
points.

A natural question to ask is whether going off shell does not break gauge
invariance, thus possibly spoiling the on-shell dual conformal symmetry. We can
answer this question in the following way. In the off-shell regime we expect two
phenomena to take place. Firstly, the double-log singularities get additional
contributions from a new subprocess associated with exchanges of particles
carrying infrared (or `ultra-soft') momenta \cite{Korchemsky2,Kuhn99}. {This
contribution is cancelled, however, in the finite part of the four-gluon amplitude defined
as the ratio of the off-shell scattering amplitude and the
off-shell form factors. At the same time, the finite part of the off-shell
amplitude receives $s/t-$dependent contributions from a hard subprocess in which
the particle momenta are of order $\sqrt{-s},\ \sqrt{-t}$. The latter is not
sensitive to the virtuality of the external legs (recall the one-loop example of
Section \ref{onelex}). We thus expect that the gauge dependence of the finite
part of the scattering amplitude may affect only the constant $s/t-$independent
term which is scheme dependent anyway~\cite{Humpert}.}

Finally, let us comment on the Wilson loop. In $\mathcal{N}=4$ SYM, the Wilson
loop evaluated along a smooth closed contour is UV finite and has conformal
symmetry.  However, the presence of cusps causes specific `cusp' UV divergences.
To regularize them in the perturbative calculation, we had to introduce a
dimensional regulator which breaks the conformal symmetry. Surprisingly enough,
the finite part of the light-like Wilson loop with four cusps on the integration
contour has exactly the same special form as the four-gluon amplitude. We are tempted to interpret
this fact as a signal that the basic reason why the finite part is always the
same is conformal symmetry. It appears to be broken in a controlled way in all
three calculations (the perturbative on-shell gluon amplitude of Bern {\it et al}, its
strong coupling dual of Alday and Maldacena, and our perturbative Wilson loop),
leaving as a common trace the specific form of the finite part.

\section*{Acknowledgments}

We have profited form enlightening discussions with L. Alday, I. Antoniadis, Z. Bern, A.
Davydychev, S. Ferrara, J. Henn, D. Kosower, L. Lipatov, J. Maldacena, V. Smirnov
and R. Stora. Special acknowledgements are due to A. Brandhuber, P. Heslop and G.
Travaglini and to L. Dixon who have shared with us their observations about the
exponentiation properties of the on-shell four-gluon amplitude. ES is very
grateful to the Department of Physics and Astronomy at UCLA and especially to Zvi
Bern for hospitality during the initial stages of this work. This research has
been partially supported by the French Agence Nationale de la Recherche, contract ANR-06-BLAN-0142.

\end{document}